\documentclass[%
 reprint,
 amsmath,amssymb,
 aps,
]{revtex4-1}

\usepackage{graphicx}
\usepackage{dcolumn}
\usepackage{bm}
\usepackage{wasysym}

\begin{document}

\preprint{APS/123-QED}

\title{
Revealing Correlated Electron-Nuclear Dynamics in Molecules \\
with Energy-Resolved Population Image
}

\author{Kunlong Liu, Pengfei Lan,$^{\dag}$ Cheng Huang, Qingbin Zhang, and Peixiang Lu}

\email{lupeixiang@hust.edu.cn}
\email{$^{\dag}$ pengfeilan@hust.edu.cn}

\affiliation{
School of Physics, Huazhong University of Science and Technology, Wuhan 430074, China
}%

\date{\today}

\begin{abstract}
We explore a new fashion, named \textit{energy-resolved population image} (EPI), to represent on an equal footing the temporary electronic transition and nuclear motion during laser-molecular interaction.
By using the EPI we have intuitively demonstrated the population transfer in vibrational H$_2^+$ exposed to extreme ultraviolet pulses, revealing
the energy sharing rule for the correlated electron and  nuclei.
We further show that the EPI can be extended to uncover the origins of the distinct energy sharing mechanisms in multi-photon and tunneling regimes.
The present study has clarified a long-standing issue about the dissociative ionization of H$_2^+$ and
paves the way to identify instantaneous molecular dynamics in strong fields.
\begin{description}
\item[PACS numbers]
33.80.Rv, 42.50.Hz, 33.80.Wz
\end{description}
\end{abstract}

\pacs{Valid PACS appear here}
\maketitle

An inner knowledge of electronic and nuclear dynamics in laser-molecular interaction has been highlighted due to the potential applications that rely on this information.
The applications include high-order harmonic generation \cite{molehhg1,molehhg2}, nuclear dynamics detection \cite{sbaker,mlein}, self-imaging of molecules \cite{cdlin}, attosecond control of formation and rupture of chemical bonds \cite{Roudnev,Kling}, and so on.
Despite more than 20 years of intensive research in this field \cite{HH2004},
the correlated electron-nuclear dynamics in molecules still attracts continuous interest inspired by the unexpected and even counterintuitive phenomena in laser-driven molecular fragmentation \cite{EI2,EI1,esry,Staudte,He,mioni}.

In the past, the underlying molecular dynamics has been typically treated as a ``black box'', of which the immediately apparent characteristics are hidden from direct observation, and researchers could only speculate the underlying mechanism according to the measurable inputs (laser parameters) and the observable outputs (electron/ion momentum distribution, high harmonics spectrum \textit{et al}.). However, such methodology is defective because, potentially, distinct dynamical models could be referred from similar observed phenomena. For instance, two distinct mechanisms, above threshold Coulomb explosion \cite{esry} and charge-resonance enhanced ionization \cite{Staudte}, were proposed to be responsible for
the modulation of the nuclear kinetic energy release (KER) spectrum following ionization of H$_2^+$, leading to a long-standing controversy \cite{esry,Staudte,contro1,contro2,contro3}.

Recently, joint energy spectrum (JES) is proposed to study the correlated electron-nuclear dynamics in dissociative ionization of molecules \cite{Silva,JES,Fischer,wuj}.
However,
it is still unclear how much of the photon energy is respectively deposited to the fragments \textit{during} the interaction.
It is even found that the way for electron and nuclei to share energy in tunnel ionization differs from that in multi-photon regime \cite{Silva}, but the origins of the distinct energy sharing mechanisms is somewhat hidden.
Since the Born-Oppenheimer (BO) approximation fails to treat the correlated electron-nuclear dynamics in molecules \cite{JES}, an approach that reveals on an equal footing the instantaneous electronic and nuclear dynamics is desired.

On one hand, extracting the instantaneous correlated electron-nuclear dynamical information from the black box of laser-molecular interaction requires solving the time-dependent Schr\"{o}dinger equation (TDSE)
beyond the BO approximation \cite{H2j,soft}.
However,
the wavepackets in space representation can hardly provide intuitive insights into the instantaneous dynamics.
On the other hand, the diagram of molecular potential curves was widely used to understand the molecular dynamics, but the qualitative population transfer usually involves speculations based on the observables.
Therefore,
presenting quantitatively the population transfer upon the potentials
through TDSE solutions
would be a more reliable and transparent manner to demonstrate the evolution of the molecular dynamics.
In this paper, we
accomplish this by introducing an intuitive representation,
named energy-resolved population image (EPI),
on the basis of a resolvent method \cite{schafer} that allows for extracting the population of not only the continuum but also the bound states.
By using the EPI, we have theoretically studied the dissociative ionization of vibrational H$_2^+$ exposed to extreme ultraviolet (XUV) pulses.
By demonstrating the temporary population transfer,
we have deduced the rule for respective amounts of the energy taken by the correlated electron and nuclei.
We further demonstrate the dissociative ionization in multi-photon and tunneling regimes with the EPIs, intuitively revealing the origins of the distinct energy sharing mechanisms.

For numerical simulations we
solved the TDSE for a reduced-dimensionality model of H$_2^+$. In this model, the one-dimensional motions of the nuclei and the electron are assumed to remain aligned with the linearly polarized laser field.
Even so, this model was widely used to identify the strong-field processes \cite{HH2004,mioni,JES,Silva} and reproduced experimental result at least qualitatively \cite{soft,Staudte}.
Thus, the simplified model of H$_2^+$ is practical and reliable to study molecular dynamics in strong field.
Within this model, the length gauge TDSE can be written as (atomic units are used throughout)
$i\frac{\partial}{\partial t}\Psi(R,z;t)=[T_N+T_e+V_0+V_t]\Psi(R,z;t)$,
where
$T_N=-\frac{1}{m_p}\frac{\partial ^2}{\partial R^2}$, $T_e=-\frac{1}{2}\frac{\partial ^2}{\partial z^2}$,
$V_t=\varepsilon(t)z$, and
$V_0=\frac{1}{R}
+V_e(z,R)$
with $V_e(z,R)$ the improved soft-core potential that reproduces the exact $1s\sigma_g$ potential curve in full dimension \cite{soft}.
Here, $R$ is the internuclear distance, $z$ is the electron position measured from the center-of-mass of the protons, and $m_p$ is the mass of the proton.
The laser electric field is given by $\varepsilon(t)=\varepsilon_0\exp[-2\ln2(t/\tau)^2]\sin(\omega t)$ with $\tau$ the pulse duration, $\omega$ the central frequency, and $\varepsilon_0$ the peak electric field amplitude.

\begin{figure}[t]
\centering\includegraphics[width=8.5cm]{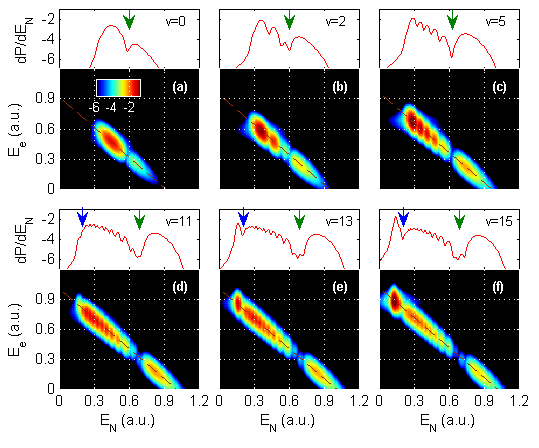}
\caption{\label{fig1}
The JES along with the KER spectra (top panels) for the interaction of the vibrational H$_2^+$ with a 0.8-fs XUV pulse at 30 nm and a peak intensity of $10^{14}$ W/cm$^2$.
}
\end{figure}

The TDSE is solved on a grid by using the Crank-Nicolson split-operator method with a time step of $\Delta t=0.04$ a.u.. The grid ranges from 0 to 25 a.u. for $R$ and from $-1500$ to 1500 a.u. for $z$, with grid spacings of $\Delta R=0.05$ a.u. and $\Delta z=0.2$ a.u..
To obtain intuitive insights into the temporary molecular dynamics, we now introduce the EPI.
The EPI is analogous to those diagrams with molecular potential curves and sketched wavepacket profiles on them, but provides a quantitative and accurate description of the population transfer.
The general idea of calculating the EPI is to convert the wave function $\Psi(R,z;t)$ at $t$, or $\Psi_t(R,z)$, to the density distribution $\rho(R,E)$ with $E$ the potential energy.
We have accomplished the conversion by extracting the energy density distribution from the $z$-dimensional wave function of $\Psi_t(R,z)$ at each internuclear distance. The extraction is based on the resolvent method, which was first introduced by Schafer and Kulander \cite{schafer}. Briefly, an energy window operator is defined by
$\hat{W}_t(E,k,\epsilon)=\epsilon^{2k}/[(\hat{H}_t-E)^{2k}+\epsilon^{2k}]$,
with $\hat{H}_t=T_e+V_0+V_t$. The probability density of the energy $E$ at each $R$ can be obtained from
$\rho(E;R)=\langle\Psi_t(z;R)|\hat{W}_t|\Psi_t(z;R)\rangle/C$
with $C=\epsilon\frac{\pi}{k}\csc(\frac{\pi}{2k})$ \cite{Silva}. Here we use the parameters of $\epsilon=0.004$ and $k=2$.

\begin{figure}[t]
\centering\includegraphics[width=8.5cm]{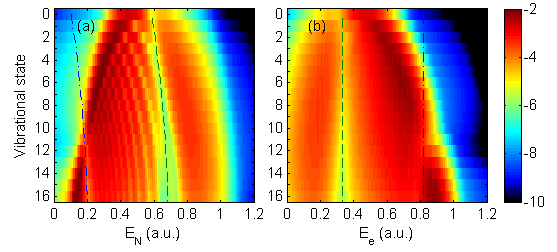}
\caption{\label{fig2}
The kinetic energy spectra of the nuclei [Panel (a)] and the electron [Panel (b)] for vibrational states from $v=0$ to 16. The dashed and dash-dotted lines indicate the locations of the suppression.
}
\end{figure}

First of all, Figure~\ref{fig1} shows the JES for the interaction of H$_2^+$ with a 0.8-fs XUV pulse at 30 nm and a peak intensity of $10^{14}$ W/cm$^2$. The JES are obtained by using the method from \cite{Silva}. A number of initial vibrational states ($v=0$, 2, 5, 11, 13 and 15) of H$_2^+$($1s\sigma_g$) have been chosen.
From Fig.~\ref{fig1}, two main spectral features are found. Firstly,
as indicated by the dashed lines, the maxima of the JES is along the lines given by
\begin{eqnarray}
E_N+E_e=E_{sys}(v)+\omega
\end{eqnarray}
with $E_{sys}(v)$ the bound energy of the $v$-th vibrational state.
This feature is governed by the energy conservation.
Secondly, the density distributions in the JES are modulated and also suppressed at some locations.
By integrating the JES over $E_e$, we show the KER spectra in the top panels of Fig.~\ref{fig1}.
More peaks appear in the spectra of higher vibrational states. The
suppressions indicated by the arrows are also observed.
Furthermore, in Fig.~\ref{fig2} we respectively show the kinetic energy spectra of the nuclei and the electron for vibrational states from $v=0$ to 16.
It is found that
the suppression locations in KER spectra shift to higher energy with increasing $v$ while those in electronic energy spectra stay constant.

\begin{figure}[t]
\centering\includegraphics[width=8.5cm]{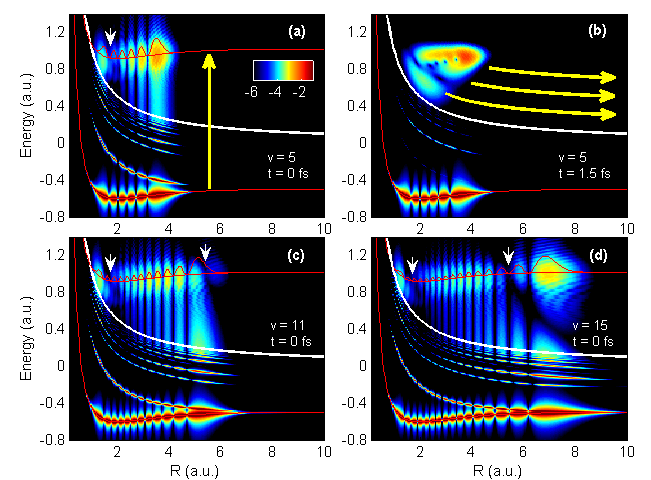}
\caption{\label{fig3}
The EPIs during the interaction. The vibrational state and the time for each EPI have been given in the corresponding panel. The thick red and white curves indicate the $1s\sigma_g$ potential and the $1/R$ curves. The thin red curves denote the up-shifted potential of $1s\sigma_g+\omega$ and the profile of the initial nuclear wavepacket distributions.
}
\end{figure}

To find out the origins of the progression of the JES,
we have simulated the evolution of the EPI for the interaction of H$_2^+$ ($1s\sigma_g$, $v=5$) with the XUV pulse. Each frame of the evolution is calculated at every half optical cycle when $\varepsilon(t)=0$.
With the time-dependent EPIs (see the multimedia),
we have quantitatively visualized the population transfer
during the interaction.
In Figs.~\ref{fig3}(a) and 3(b), we show the EPIs at $t=0$ and 1.5 fs, respectively. For reference,
the $1s\sigma_g$ potential $V_g(R)$ (the thick red curve), the Coulomb explosion curve (the thick white one) and the profile of the initial nuclear wavepackets on the up-shifted potential curve $V_g(R)+\omega$ (the thin red one) are plotted on the EPI.
As indicated by the thick arrows,
the system firstly absorbs one photon energy and
the population is transferred to the continuum with maxima along the $V_g(R)+\omega$ curve.
Then,
the outgoing population at each $R$ slides down along respective routes parallel to the $1/R$ curve.

\begin{figure}[t]
\centering\includegraphics[width=8.5cm]{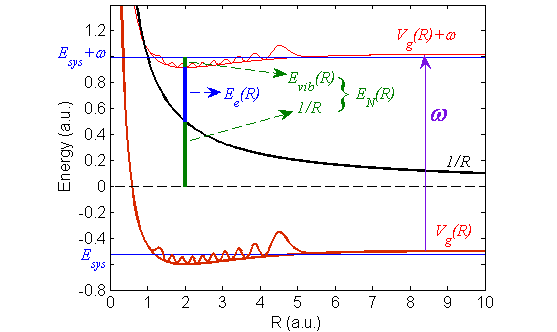}
\caption{\label{fig4}
Illustration of electron-nuclear energy sharing in single-photon induced dissociative ionization of H$_2^+$.
}
\end{figure}

Based on the EPIs,
we now deduce the electron-nuclear energy sharing rule.
As shown in Fig.~\ref{fig4},
via absorbing a photon energy of $\omega$, the population is transferred to the position given by $E(R)=V_{g}(R)+\omega$.
For the electron, the ponderomotive energy is close to zero due to the ultrashort wavelength of the XUV pulse. The electronic energy will stay constant after the transfer and thus
the final electronic kinetic energy is
\begin{eqnarray}
E_e(R)=V_{g}(R)+\omega-1/R,
\end{eqnarray}
where $1/R$ indicates the ionization threshold at $R$.
For the nuclei,
according to Eqs. (1) and (2),
the nuclear kinetic energy reads
\begin{eqnarray}
E_N(R) =E_{vib}(R)+1/R
\end{eqnarray}
with $E_{vib}(R)=E_{sys}(v)-V_{g}(R)$ the vibrational energy of the nuclei when Coulomb explosion starts, indicating that the final nuclear kinetic energy includes the initial vibrational energy and the energy that arises from Coulomb explosion.

The EPI in Fig.~\ref{fig3}(a) also demonstrates
that the population being transferred to the continuum
is proportional to the bound population and is suppressed at the internuclear distance indicated by the small arrow.
This has been confirmed via the EPIs for other vibrational states shown in Figs.~\ref{fig3}(c) and~\ref{fig3}(d).
The yields of the correlated electron (with energy $E_e$) and nuclei (with energy $E_N$) are thus
\begin{eqnarray}
Y[E_N(R),E_e(R)]\propto |\chi(R)|^2\cdot\Gamma(R)
\end{eqnarray}
with $\chi_v(R)$ the nuclear wave function of the $v$-th vibrational state and $\Gamma(R)$ the ionization rate at $R$.
Generally, the energy sharing rule given by Eqs. (2) and (3), together with Eq. (4), indicates
that the yields of the correlated fragments at different energies are (i) determined by the $R$-distributed population and (ii) affected by the ionization rates at different $R$.

With Eqs. (2)--(4),
now the spectral features in Figs. 1 and 2 can be well understood.
Because there are more peaks in the nuclear wavepacket distribution of higher vibrational states,
more maxima appeared in the KER spectra and the JES.
The peaks in electronic energy spectra are blurred due to the broadband photon energy of $\omega$ in Eq. (2).
On the other hand, according to the previous studies on the electronic wavepacket interference \cite{huang,Henkel},
the ionization would be suppressed at critical internuclear distances due to the destructive interference of the ionized electronic wavepackets from two nuclei.
As shown in Fig. 3, the population transfer is mainly suppressed at $R_s\approx1.70$ and 5.45 a.u..
Then the locations of the suppressions can be obtained by inserting $R_s$ to Eqs. (2) and (3), respectively.
Clearly,
for a given $R_s$,
$E_{N}(R_s)$ will increase with the bound energy of $E_{sys}(v)$ while $E_{e}(R_s)$ is independent on vibrational states.
We have shown the locations given by $E_{N}(R_s)$ and $E_{e}(R_s)$ with the dashed ($R_s\approx1.70$) and dash-dotted ($R_s\approx5.45$) lines in Fig. 2, coming out
consistent with the spectral tendency.

\begin{figure}[t]
\centering\includegraphics[width=8.5cm]{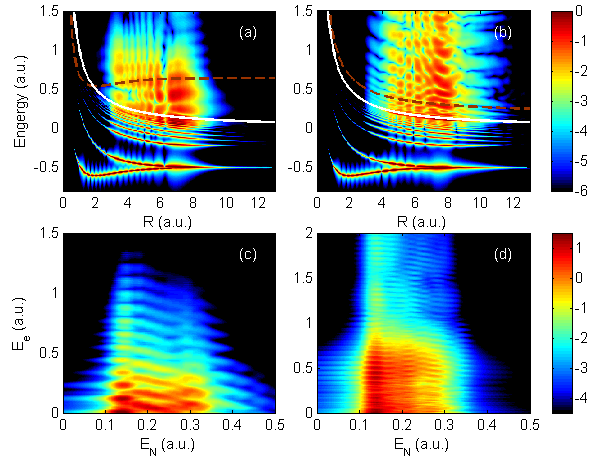}
\caption{\label{fig5}
The EPIs (upper panels) at $t=0$ and the JES (lower panels) for the interactions of H$_2^+$ ($1s\sigma_g$, $v=15$) with 400 (left column) and 800 nm (right column) pulses. The pulse duration is 3 optical cycle and the intensity is $10^{14}$ W/cm$^2$.
}
\end{figure}

Basically, the physical picture for the single-photon ionization of H$_2^+$ demonstrated by the EPI could generalize to multi-photon and tunneling regimes.
In Fig.~\ref{fig5}, we show the EPIs (upper row) for the interaction of H$_2^+$ ($1s\sigma_g$, $v=15$) with the 400 and 800 nm pulses at $t=0$ and the corresponding JES (lower row). The pulse duration is three optical cycle and the intensity is $10^{14}$ W/cm$^2$.
The values of the Keldysh parameter for Figs. 5(a) and 5(b) are 2.52 and 1.26, respectively, corresponding to the ionization processes in multi-photon and close to tunneling regimes \cite{Eckle}.
Here, we double the calculation grid
to ensure no significant reflection of
the wavepackets at the boundary during the interaction.

The EPIs in Figs. 5(a) and 5(b) demonstrate the coexistence
of two somewhat controversial mechanisms: above threshold Coulomb explosion \cite{esry} and charge-resonance enhanced ionization \cite{Staudte}.
Firstly, the \textit{vertical} arrangement of the maxima in the continuum are observed, intuitively verifying the multi-photon absorption process.
Meanwhile, strong coupling between the $1s\sigma_g$ and $2p\sigma_u$ states is observed in the range of $3<R<8$ a.u., where the continuum population is much more pronounced than that of small internuclear distance. The modulated population in bound states thus leads to the near \textit{horizontal} arrangement of the maxima in the continuum.
According to Fig. 4 and the energy sharing rule given by Eqs. (2) and (3), the maxima of each vertical column in the continuum will contribute to
the multi-peak structure of the electronic energy spectrum while the near horizontal arrangement of the maxima is responsible for the modulated structure of the KER spectrum.

Furthermore, as illustrated by the dashed lines in Figs. 5(a) and 5(b), there are distinct horizontal arrangements of the maxima in the continuum, which are associated with the ionization processes.
In multi-photon regime [Fig. 5(a)], the bound population is `vertically' transferred to the continuum via multi-photon absorption \cite{ionIva}, resulting in the population maxima along the up-shifted $1s\sigma_g$ potential. Thus, similarly to Eqs. (2) and (3), the energy sharing in multi-photon regime can be sketchily given by
$E_e'(R)\approx n\omega'+V_{g}(R)-1/R$
and $E_N'(R) \approx1/R+E_{vib}'(R)$,
which indicate that $E_e'$ and $E_N'$ are correlated through the parameter $R$, resulting in the tilted spectral structure in Fig. 5(c).
In tunneling regime [Fig. 5(b)], the bound population would first `tunnel' to the continuum and then absorb energy from the field \cite{ionIva}. As a result, the population maxima are almost along the lines parallel to $1/R$ curve. Therefore, the energy sharing in tunnel regime can be sketchily given by
$E_e''(R)\approx m\omega''$
and
$E_N''(R) \approx 1/R+E_{vib}''(R)$,
where $E_e''$ shows independence on the parameter $R$ and thus loses the correlation with $E_N''$, as the JES shown in Fig. 5(d).

In summary, an intuitive representation, i.e. the EPI, was introduced to make the instantaneous molecular dynamics visible in a quantitative way.
The population transfer and the detailed energy sharing processes during the dissociative ionization of H$_2^+$ have been intuitively demonstrated by the EPIs.
It confirms (i) that
the electron-nuclear energy sharing is
determined by the internuclear distance, nuclear vibrational energy, and photon energy, and (ii) that the yields of the correlated fragments are associated with the nuclear wavepacket distribution and $R$-dependent ionization rates.
Moreover, on the basis of the EPIs,
the different energy sharing mechanisms in multi-photon and tunnel ionization of H$_2^+$ are found to originate from the distinct ways in which the bound population is transferred to the continuum.
We emphasize that the EPI allows scientists directly ``see'' how the molecular fragmentation proceeds. It suggests a new access to study the dynamics of molecules and would be conductive to explore the underlying molecular mechanisms in future experimental studies.

This work was supported by the National Natural Sci-
ence Foundation of China under Grant No. 61275126,
11234004, and the 973 Program of China under Grant
No. 2011CB808103.

\end{document}